\documentstyle[preprint,aps]{revtex}
\begin{document}
\title{The Evolution of Nuclear Multifragmentation in the
Temperature-Density Plane}
\author{P. G. Warren$^{(1)}$, S. Albergo$^{(3)}$, J. M.
Alexander$^{(7)}$, F. Bieser$^{(6)}$, F. P. Brady$^{(2)}$,\\ 
Z. Caccia$^{(3)}$, D. A. Cebra$^{(2)}$, A. D. Chacon$^{(8)}$, J. L. 
Chance$^{(2)}$, Y. Choi$^{(1)}$,\\ 
S. Costa$^{(3)}$, J. B. Elliott$^{(1)}$, M. L. Gilkes$^{(1)}$, J. A.
Hauger$^{(1)}$, A. S. Hirsch$^{(1)}$,\\ 
E. L. Hjort$^{(1)}$, A. Insolia$^{(3)}$, M. Justice$^{(5)}$, D.
Keane$^{(5)}$, 
J. C. Kintner$^{(2)}$,\\ 
R. Lacey$^{(7)}$, J. Lauret$^{(7)}$, V. Lindenstruth$^{(4)}$, M. A.
Lisa$^{(6)}$, 
H. S. Matis$^{(6)}$,\\ 
R. L. McGrath$^{(7)}$, M. McMahan$^{(6)}$, C. McParland$^{(6)}$,
W. F. J. 
M\"{u}ller$^{(4)}$, D. L. Olson$^{(6)}$,\\ 
M. D. Partlan$^{(2)}$, N. T. Porile$^{(1)}$, R. Potenza$^{(3)}$, G.
Rai$^{(6)}$, 
J. Rasmussen$^{(6)}$, H. G. Ritter$^{(6)}$,\\ 
J. Romanski$^{(3)}$, J. L. Romero$^{(2)}$, G. V. Russo$^{(3)}$, H. 
Sann$^{(4)}$, R. P. Scharenberg$^{(1)}$,\\ 
A. Scott$^{(5)}$, Y. Shao$^{(5)}$, B. K. Srivastava$^{(1)}$, T. J. M.
Symons$^{(6)}$, M. L. Tincknell$^{(1)}$,\\ 
C. Tuv\'{e}$^{(3)}$, S. Wang$^{(5)}$, H. H.
Wieman$^{(6)}$, T. Wienold$^{(6)}$, K. Wolf$^{(8)}$\\
(EOS Collaboration)}
\address{$^{(1)}$Purdue University, West Lafayette, IN 47907\\
$^{(2)}$University of California, Davis, CA 95616\\
$^{(3)}$Universit\'{a} di Catania and Instituto Nazionale di Fisica
Nucleare-Sezione di Catania,\\
95129 Catania, Italy\\
$^{(4)}$GSI, D-64220 Darmstadt, Germany\\
$^{(5)}$Kent State University, Kent, OH 44242\\
$^{(6)}$Nuclear Science Division, Lawrence Berkeley National
Laboratory, Berkeley, CA 94720\\
$^{(7)}$State University of New York at Stony Brook, Stony Brook, New
York 11794\\
$^{(8)}$Texas A\&M University, College Station, TX  77843}
\date{\today}
\maketitle
\newpage
\begin{abstract}
The mean transverse kinetic energies, $\langle KE_{\perp}\rangle$, of
fragments
formed in the interaction of 1 A GeV Au+C have been determined.\ \
An energy balance argument indicates the presence of a collective energy 
which increases in magnitude with increasing multiplicity and accounts for 
nearly half of the measured $\langle KE_{\perp}\rangle$.\ \
The radial flow velocity associated with the collective energy
yields estimates for the time required to expand to the freeze-out volume.\ \ 
Isentropic trajectories in the temperature-density plane are
shown for the expansion and indicate that the system goes through the
critical region at the same multiplicities as deduced from a 
statistical analysis. Here, the expansion time is $\sim$70 fm/c.\\
PACS numbers:\ \ 25.75.+r
\end{abstract}

Recently, there has been renewed experimental 
\cite{Gilkes82,Pocho95,Mastinu96,Kwiat95} and theoretical interest
\cite{Dorso95,Campi88,Bondorf95,Bonasera94,Botvina95,Bauer88,Gross90,Gupta96} 
in the phenomenon known as nuclear multifragmentation (MF).\ \
With sufficient excitation energy, a large nucleus will disassemble into
nucleons, light fragments, and several intermediate mass fragments 
(IMF) $3 \leq Z_f\leq$ 30.\ \ 
A recent experiment \cite{Gilkes82} conducted at the Lawrence Berkeley 
Bevalac by the EOS Collaboration has provided a large sample of events 
in which the total charge 
of the projectile, 1 GeV per nucleon gold incident on a carbon target, has 
been completely reconstructed.\ \ Such events have permitted a detailed
fragment yield analysis that points strongly to the connection between MF and
critical behavior in a finite system \cite{Gilkes82}.\ \ The implication of 
this statistical analysis is that the system is substantially equilibrated at 
the time of fragment formation.\ \ Further support for equilibration comes 
from a study of MF using different beam energies and projectile-target
combinations \cite{Kreutz93}.\ \ In a recent paper \cite{Hauger96}, the notion 
of thermal and chemical equilibrium was employed to calculate an 
initial and final temperature of the system undergoing MF.\ \ 
Here the evolution of MF in the temperature-density $(T-\rho)$ plane 
is explored as a function of the total charged particle multiplicity.

It has long been a goal to understand the evolution of highly excited
nuclear systems in the $T-\rho$ plane.\ \ Theoretical efforts addressing
this
issue have tended to focus on the behavior of infinite nuclear matter 
\cite{Jaqaman83}.\ \ It is much more difficult to solve this
problem for finite size systems.\ \ 
One theoretical approach often taken is to circumvent the 
complete dynamical description and use statistical methods to form the 
fragments \cite{Bondorf95,Gross90}.\ \ Such methods 
describe the statistical aspects of MF reasonably well, but fail, as shown
below, to describe the dynamics satisfactorily.\ \ An alternative method is
to follow the dynamical process from the beginning.\ \ These approaches, 
however, do not adequately describe the statistical aspects 
\cite{Bertsch88}.\ \ With simple assumptions and the experimental data,
another approach is presented which describes the dynamical evolution 
of the fragmenting system.

The experimental apparatus was described by Gilkes et al \cite{Gilkes82}.\
\ 
Two essential features of the experiment, reverse kinematics and the large
mass
asymmetry between target and projectile, permit an unambiguous
identification of the source of all projectile fragments of charge $Z > 2$.
Particles with $Z \leq 2$ can be separated into two categories, those
associated
with the heavier nuclear fragments, which are emitted in the second stage
of the reaction, and those emitted in the primary collision stage.\ \ 
The source of the
heavier fragments is termed the remnant and travels at close to beam
velocity.\ \ In a recent paper \cite{Hauger96}, the remnant mass and excitation
energy were determined as a function of the measured total
charged particle multiplicity, $m$.\ \ To convert the initial excitation
energy of the remnant to an initial temperature, $T_i$, the state of 
the remnant was
assumed to be that of an equilibrated Fermi gas at a density, $\rho_i$, 
determined by the remnant mass and the unexpanded gold nuclear volume, $V_i$.
\ \ A freeze-out temperature, $T_f$,  was determined using a double ratio of
observed isotopic yields \cite{Albergo85}.

Figure 1 shows $\langle KE_{\perp}\rangle$, where $KE_{\perp}=p_{\perp}^2 /2M$,
for fragments of charge 3-6 for 5 different multiplicity intervals.\ \ 
Individual fragments had $KE_{\perp}$ values which ranged up to 100 MeV. \ \ 
Only statistical errors are shown. Systematic errors are discussed in
reference \cite{Hauger96b}. A significant
evolution has occurred in going from low $m$ to high $m$ events.\ \ For
comparison, the predictions of the intranuclear cascade calculation ISABEL 
\cite{Yariv81} coupled to the statistical multifragmentation model (SMM) 
\cite{Bondorf95} are included in Fig. 1.\ \ No radial
flow was included in the SMM-generated data.\ \ These two calculations  
model the first and second stages of the reaction, respectively.\ \ 
Substantial differences between
data and this model develop as a function of increasing multiplicity.\ \ For
the lowest $m$ interval, the simulation is in reasonable agreement with the
data, both in magnitude and as a function of charge.\ \ However, as $m$
increases, the simulation predicts that a given fragment's average transverse
kinetic energy decreases, while the data show an increase in
$\langle KE_{\perp}\rangle$ with $m$.\ \ Similar trends are also found for 
the heavier fragments.\ \ 

The behavior in SMM can be understood as follows.\ \ 
SMM produces spectra by randomly placing
fragments in a volume that increases with multiplicity.\ \ Fragments are
given thermal energy and are then evolved classically under the
influence of the Coulomb force.\ \ The decrease in a fragment's kinetic
energy in the simulation with increasing $m$ is due to both the increase
in volume and the decrease in the charge of the largest fragment.\ \ The
average charge of the largest fragment as a function of $m$ in data and
SMM are in very good agreement, so this is not the source of the
discrepancy.\ \ Likewise, no reasonable adjustment of the distance between
fragments can lead to resolution of the discrepancy.\ \ The 
excitation energy and mass spectra of the remnant predicted by ISABEL
are also 
in reasonable agreement with the data \cite{Srivastava}.

It will be assumed that the initial hot remnant undergoes expansion 
to freeze-out with no decrease in nucleon number. 
Thus, at freeze-out, not only the IMFs, but all particles associated with the 
breakup of the remnant must be taken into account. Energy conservation relates 
the initial excitation energy of the remnant, $E_i^*$, to the energy 
at freeze-out.\ \ 
This latter energy can be decomposed into four contributions:
\begin{equation}
E_i^*~=~E_{th} + E_C + Q_f + E_X~,
\end{equation}
where $E_{th}$ is the total thermal contribution to the spectra, $E_C$ is
the Coulomb contribution, $Q_f$ is the sum of the Q-value contribution for 
each final
state particle, and $E_X$ is what remains when the other three terms are
subtracted from $E_i^*$.\ \ Within the spirit of the calculation of the final
state temperature, each final state particle originating in the breakup of the
remnant will contribute ${3 \over 2} T_f$ to $E_{th}$.\ \ The total
Coulomb energy available for doing work is given by the difference in the
self-energy of the remnant and the sum of the self-energies of all the
second stage particles.\ \ Thus,
\begin{equation}
E_C~=~{3 \over 5}~e^2 \left({Z_R^2 \over R_{Au}}~-\sum~{Z_f^2
\over R_f}\right)~.
\end{equation}
In Eq. 1, the contribution of the unobserved neutrons are estimated as 
in \cite{Hauger96}.

With Eq. 1, $E_X$ can be calculated for the experimental data.\ \ 
This quantity, displayed in Fig. 2 as a function of $m$, increases from
1 to 6 MeV/nucleon with increasing $m$, thus accounting for nearly
half of $\langle KE_{\perp}\rangle$ that is not explained 
by the thermal, Coulomb, and Q-value terms of Eq. 1.\ \ 
For the simulated data, a freeze-out temperature
was determined using the Albergo procedure \cite{Albergo85}.\
 \ This freeze-out temperature is in good agreement with that obtained from 
the experimental data \cite{Srivastava}.\ \ When the same 
energy conservation is applied to the ISABEL-SMM
generated spectra, a value of $E_X$ of approximately zero is obtained.\ \ 
The nature of the energy, $E_X$, in the data is now addressed.

Two characteristics of the kinetic energy spectra of IMFs have long been 
noted: the Coulomb energy contribution is much
smaller than that estimated assuming surface emission from a normal
density remnant system, and the inverse slope parameter characterizing the high
energy tail of the spectra is much larger than the thermal energy inferred 
from isotopic yields \cite{Poskanzer71,Hirsch84}.\ \ The low 
Coulomb energies support the view that fragment formation occurs in an 
expanded system \cite{Friedman90,Kwiat94}. However, there is no general
consensus on the interpretation of the kinetic energy spectra  
\cite{Hirsch84,Barz89,Boal89}.

A possible resolution to the discrepancy between the freeze-out temperature 
and the $\langle KE \rangle$ may lie in the quantity $E_X$.\ \ 
It will be assumed that entropy is generated during the rapid cascade stage 
of the collision process and that thereafter the remnant expands at constant
entropy \cite{Sobel75,Bertsch81}.\ \ The quantity $E_X$ may then represent 
the collective motion generated during the expansion \cite{Cugnon84,Aguiar92}.
\ \ Statistical models such as SMM must add collective energy explicitly since 
they do not perform a true dynamical evolution of
the remnant system.\ \ The addition to SMM of collective energy of the
same magnitude as indicated in Fig. 2 substantially diminishes the
discrepancies of the trends and magnitudes between the simulation 
and the data \cite{Warren96}.

If the remnant expands isentropically, 
the freeze-out volume can be estimated 
using the initial temperature and volume of the remnant, $T_i$ and $V_i$, 
and the freeze-out temperature, $T_f$, \cite{Hauger96}.
For a non-dissipative Fermi gas, $V_f = V_i (T_i /T_f )^{3/2}$.\ \ 
Figure 3 shows the freeze-out density, obtained by 
dividing the remnant nucleon number by the freeze-out volume.\ \ 
This yields a density about 1/3 normal nuclear matter in the 
multiplicity interval identified as the critical region by a statistical 
analysis of this data \cite{Gilkes82}.

The assumption of an isentropic expansion can be checked by 
estimating the entropy per particle before and after expansion of the remnant.\
\ The initial entropy is taken to be that of a Fermi gas at the temperature 
and density of the remnant and ranges from 1 to 2.5 per particle 
as a function of $m$.\ \ 
The entropy in the expanded state is computed using the SMM prescription with 
the freeze-out temperature and volume taken to be 
$T_f$ and $V_f$, respectively.\ \ 
The final entropy is systematically 
lower than the initial entropy by about 0.2 per particle.\ \ 
However, because it was assumed that all of the initial excitation energy 
was thermalized, the initial entropy estimated represents an upper bound. 
Reducing the initial excitation energy
per nucleon by only 5\% brings the initial and final entropies into agreement.
The signficant point is that no additional entropy is generated 
in the fragmentation process, supporting the assumption of an isentropic
expansion. 

For an isotropically expanding remnant, the average time
required for the system to expand to freeze-out can be estimated 
by dividing the increase in radius by the mean flow velocity, 
$\langle v_f\rangle = v_f/2$.\ \ This estimate yields 
a short expansion time of about 70 fm/c, 
in good agreement with expansion times obtained from
fragment-fragment correlation studies \cite{Bauge93,Fox94} and
theoretical predictions \cite{Gross90}.\ \ 
The time is not strongly dependent 
on multiplicity. Thus, multifragmentation is a very
fast process which suggests a simultaneous disassembly of the remnant.
The presence
of a significant component of directed sideward flow could affect
this time estimate.\ \ However, it is unlikely that such flow is of importance
in the very asymmetric Au+C collisions.\ \ It is known that 
directed flow is 
maximal at intermediate multiplicities \cite{Lisa95} 
whereas the observed value of $E_X$ becomes largest at the highest $m$.

Fig. 4 shows the evolution of the remnant system 
from initial density and temperature to freeze-out density and temperature.\ \ 
The remnant system is driven towards lower $\rho$
and $T$ via expansion.\ \ Multiplicities in the critical region,
as deduced from a statistical analysis of the data,
are indicated by bold lines which follow the $VT$ relationship for 
an isentropic expansion of a Fermi gas.\ \ 
The region in the $T-\rho$ plane for which critical behavior is expected for a 
finite charged nucleus \cite{Jaqaman83} is about 1/3 normal nuclear density 
and approximately between 5-8 MeV in temperature, i.e. $T_c \sim E_B$, 
the binding energy/nucleon for a finite nucleus\cite{Jaqaman83,Das92}.\ \ 
This dynamical analysis indicates that as the initial excitation energy is
increased the system begins to enter the critical region for the same 
multiplicities (excitation energies) as deduced from the statistical analysis.\
\ This conclusion is based on the assumption of energy and entropy
conservation.

In summary, an examination of the behavior of the mean transverse kinetic 
energies for fragments of charge 3-6 for five different multiplicity intervals 
and an exercise in energy conservation
have shown evidence for collective motion increasing with multiplicity 
and thus, with excitation energy.\ \ 
Time scales for expansion agree well 
with both previous experimental determinations and theoretical
estimates.\ \ Isentropic trajectories in the 
neighborhood of the critical point in the temperature-density plane 
have been determined from experimental data for the first time.\ \ 
Together with the statistical analysis of the data, a
strong case can be made for thermalization in multifragmentation.

This work was supported in part by the U.S. Department of Energy under 
Contracts or Grants No. DE-AC-03-76SF00098, DE-FG02-89ER40531, 
DE-FG02-88ER40408, DE-FG02-88ER40412, DE-FG05-88ER40437, and
by the
U.S. National Science Foundation under Grant No. PHY-91-23301.

\newpage
\begin{center}
Figures
\end{center}
\noindent Figure 1.\ \ Average transverse kinetic energy as a function of
multiplicity bin for fragments of charge 3-6.\ \ Circles are data and
triangles are results of ISMM (ISABEL+SMM) simulation without the
addition of flow.\ \ Multiplicity intervals are: 1-10, 11-20, 21-30, 31-40, 
41+. 
\vspace{12pt}
\noindent\\
Figure 2.\ \ Energy per remnant nucleon versus total charged particle
multiplicity.\ \ Squares are the remnant excitation energy; circles are sum
of thermal, Coulomb and Q-value contributions.\ \ The energy $E_X$, the 
difference between the two quantities, is shown as triangles.
\vspace{12pt}
\noindent\\
Figure 3.\ \ Initial (open squares) and final (filled squares) densities as a
fraction of normal nuclear density versus $m$.
\vspace{12pt}
\noindent\\
Figure 4.\ \ Trajectories in the temperature-density plane shown for
different multiplicities.\ \ Trajectories in the neighborhood of the critical
region (as determined from the statistical analysis in [1]) are in boldface.
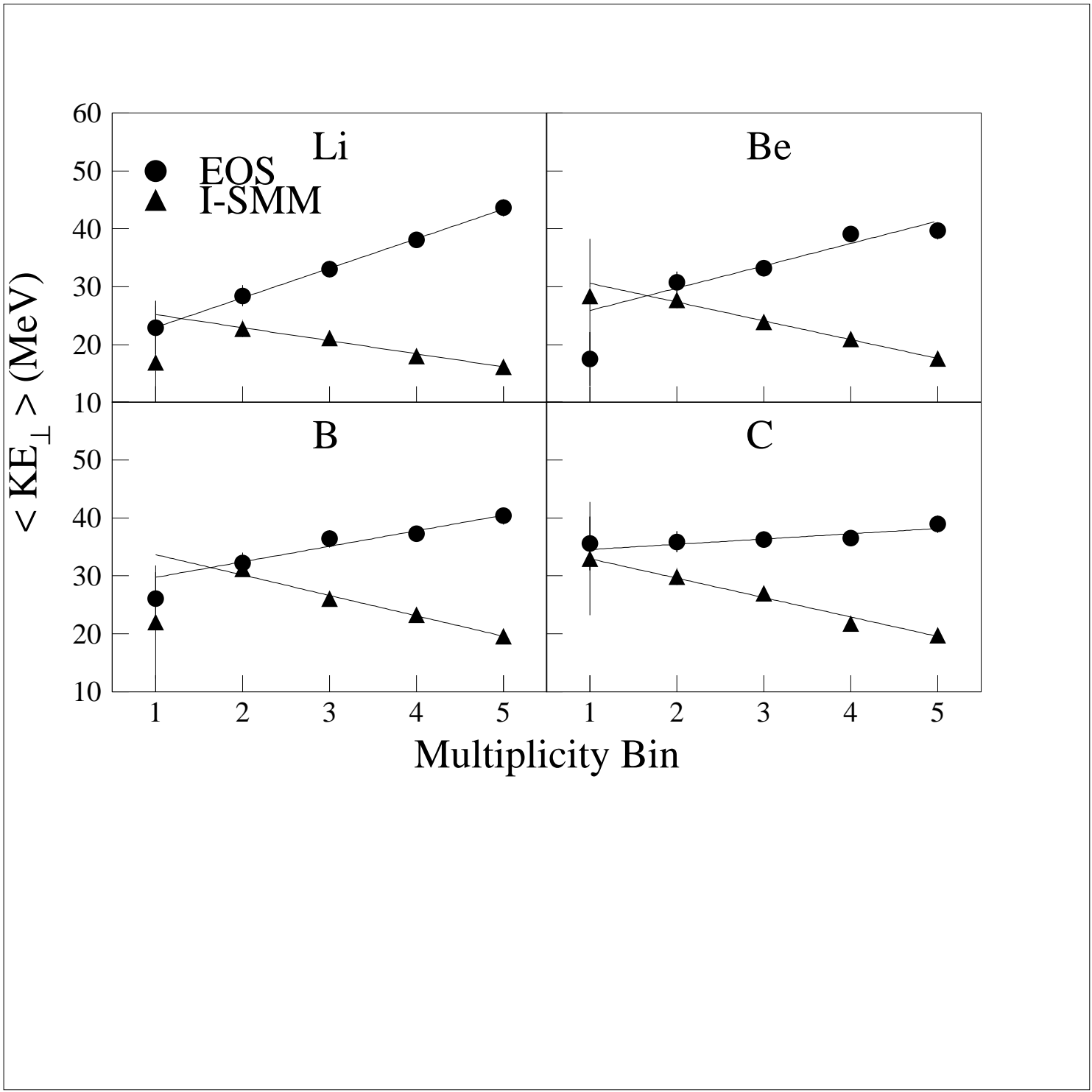
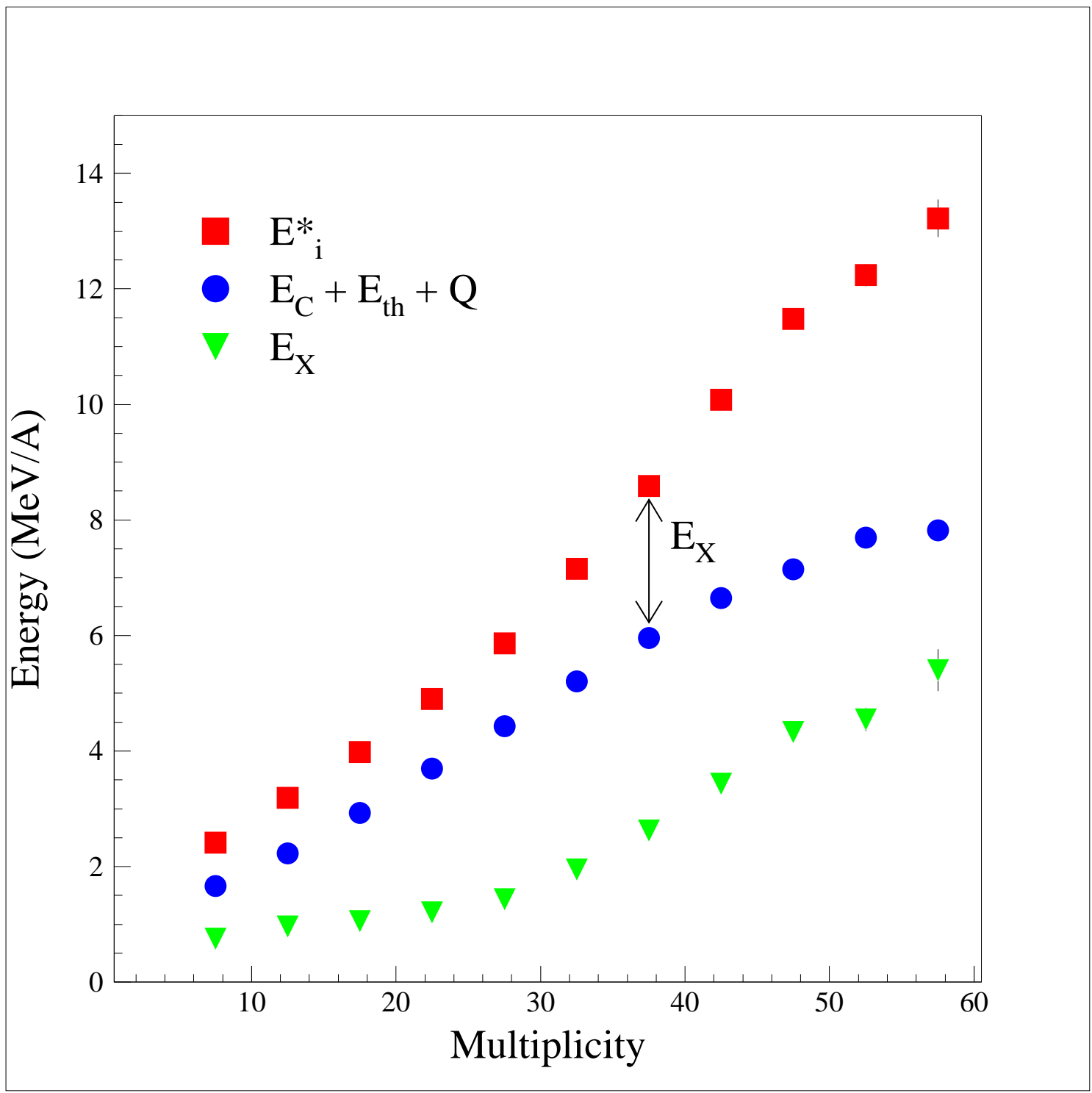
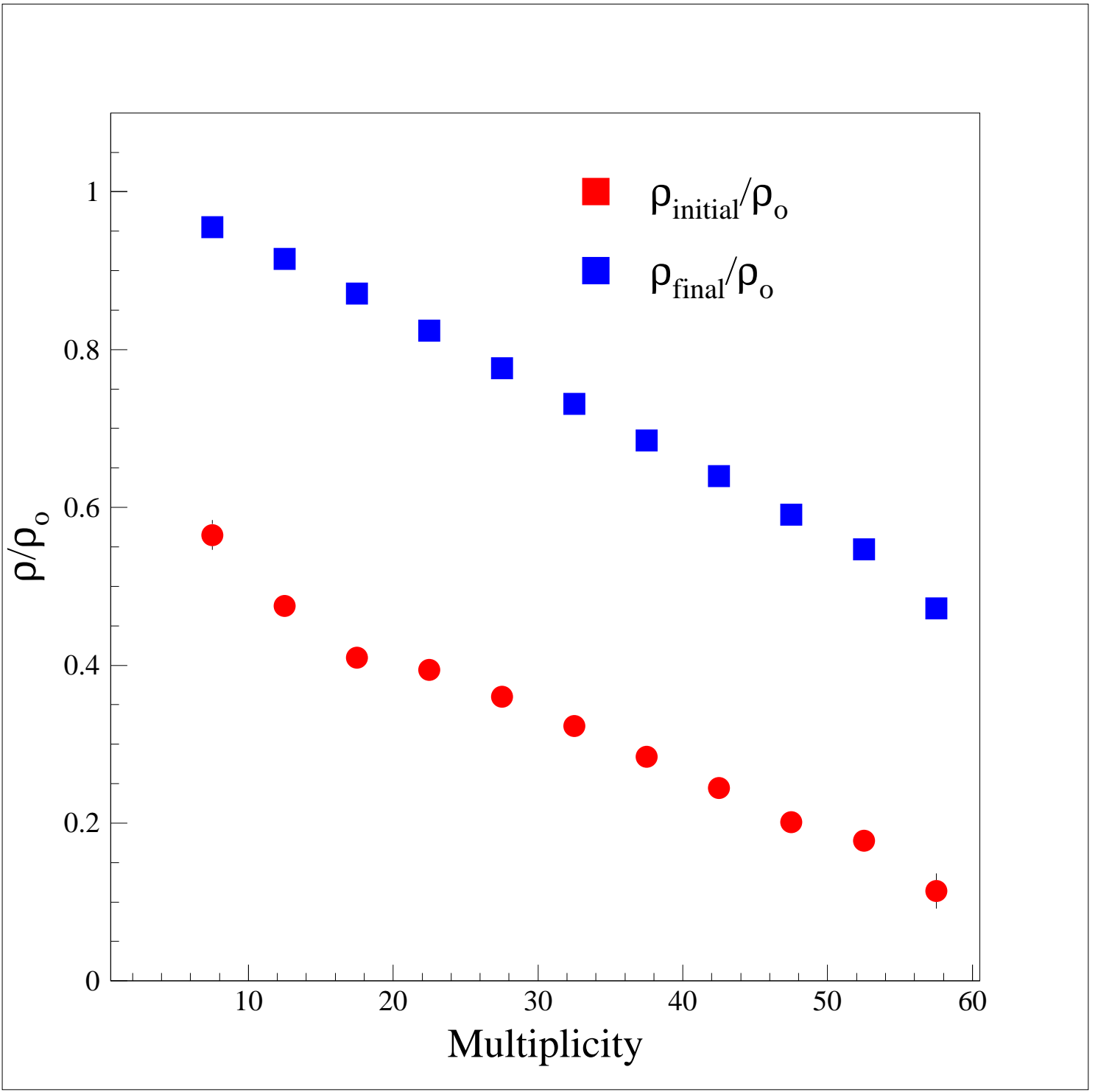
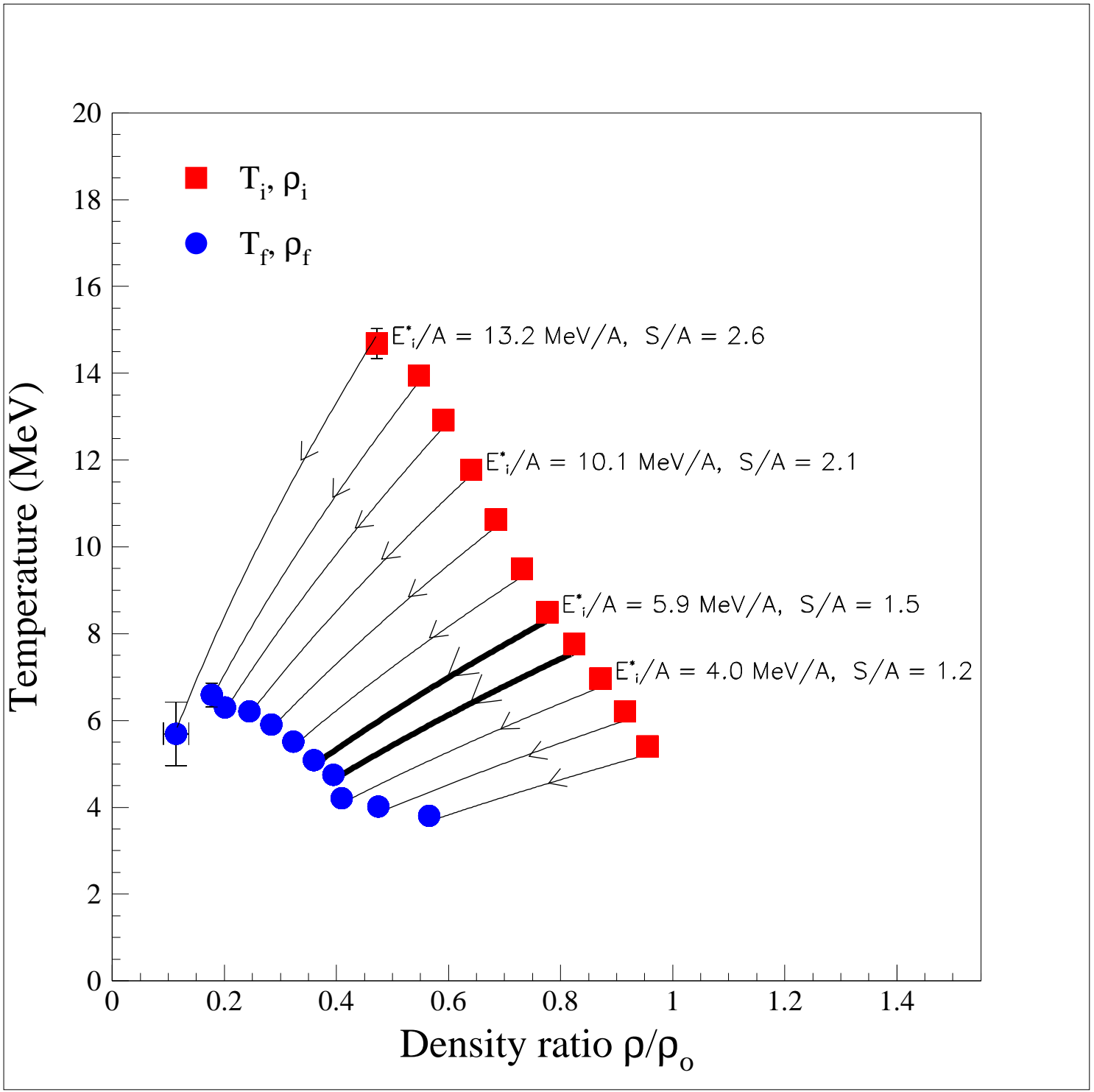
\end{document}